Francois-Igor PRIS (И. Е. Присъ)

PhD in philosophy, PhD in theoretical physics
Institute of philosophy of the National Academy of Sciences of Belarus
frigpr@gmail.com

## ON QUANTUM FUNDAMENTALISM[1]

*Abstract.* According to one possible diagnosis of the quantum measurement problem, it is a consequence of quantum fundamentalism claiming that ontology and epistemology of the world are exclusively quantum, and classical physics is only an approximation. For N. Bohr, the measurement problem is a pseudo-problem because any quantum phenomenon presupposes the classical context of an experimental setup and the use of classical concepts to describe it. We consider Bohr's position from the point of view of our contextual quantum realism (CQR), inspired by the later Wittgenstein philosophy. Our approach is consistent with H. Zinkernagel's interpretation, according to which Bohr's position is not only epistemological anti-fundamentalism, but also ontological anti-fundamentalism.

### 1. The measurement problem and Bohr's epistemological anti-fundamentalism

The quantum measurement problem has various incarnations and guises. One possible diagnosis of the problem is that it is led by quantum fundamentalism, which claims that the ontology and epistemology of the world are exclusively quantum, and that classical physics is only an approximation to quantum physics. Quantum fundamentalism is a purely physicalist approach, according to which a measurement is always seen as a quantum entanglement of a measuring instrument and an observed quantum system, which does not allow one to answer the question of how a particular observable value of a quantum physical quantity registered by the instrument arises, or even why it arises at all. The result of "measurement" as a quantum interaction/entanglement is a superposition in which the wave function of the instrument is entangled with the wave function of the observed system. Decoherence, which takes into account the interaction of the quantum system with the environment (and, in particular, with the instrument itself) having many degrees of freedom, as considered by most researchers, does not answer the question of why one and only one measurement outcome occurs either.

N. Bohr's philosophical position is not quantum fundamentalism, in the sense that, according to it, all objects can be regarded as quantum, but not simultaneously. Bohr draws an epistemological distinction between the instrument of measurement – a space-time "frame of reference", relative to which a quantum physical quantity can only have a particular value – and the object of measurement. The former is separate from the latter and receives information about it. This presupposes the use of classical concepts. That being said, for Bohr the boundary between the domain of application of classical concepts (instrument) and the domain of application of quantum concepts (quantum system) does not go along the line macroscopic/microscopic. For him, sometimes parts of a macroscopic system can belong to the domain of application of quantum concepts (quantum mechanics). It is important to note that, for Bohr, the "interaction" of the instrument and the object of observation should not be regarded as a mechanical/physical influence. The influence of the observer (instrument) manifests itself in the choice of experimental conditions of observation and, consequently, the conditions of description of the quantum system.

[1] The reported study was partially funded by BRFFR according to the research project № Г22МС-001 «Quantum realism – contextual realism. (QBism and other interpretations of quantum mechanics from the point of view of a contextual realism.)»

These conditions are described by classical concepts.[2] It is Bohr's use of classical concepts that allows him to get rid of the measurement problem as a pseudo-problem. Bohr rejects the existence of hidden variables and treatment of the wave function collapse as a real physical process (see also [1]).

According to Henrik Zinkernagel, Bohr "dissolves" the measurement problem because for him the quantum world makes sense only in relation to a certain classical description and a certain classical world [2]. In terms of our position, contextual quantum realism (CQR), this simply means that Bohr takes into account the "grammar" of a phenomenon. Any phenomenon, including quantum phenomena, has the following "classical" logic: something appears (is given) to someone ("classical" observer), at some point of ("classical") time, in some ("classical") way and under some ("classical") conditions. This structure of a phenomenon was pointed out by Aristotle.[3] Nevertheless, exactly this Bohrian notion – in our view erroneously – has been criticized by some physicists and philosophers as a harmful pseudo-notion. For example, J. A. Wheeler and, following him, W. A. Miller and C. De Ronde claimed that Bohr's main concept of the "elementary quantum phenomenon" is a "great smoky dragon", i.e. a pseudo-concept [4, p. 73; 5]).[4] The Copenhagen interpretation of Bohr is also criticized by D. Deutsch and many other physicists and philosophers of physics [6, ch. 12].

Zinkernagel argues that for Bohr, not only epistemologically, but also ontologically, the distinction between the classical and the quantum is contextual [2]. If this is so, then Bohr's position can unequivocally be interpreted within the framework of our CQR. Researchers, however, have no consensus on whether Bohr really believed that the classical world is not only epistemically but also ontologically different from the quantum world. It is important to note here that Bohr sees a certain analogy between the special theory of relativity (STR) and quantum mechanics. For him, different experimental contexts of observation in quantum mechanics play the role of different frames of reference in STR. According to M. Dorato, this means that Bohr regarded quantum mechanics as a principle theory and this explains both the epistemic dependence on the field of classical physics, which Bohr introduces, and his ban on any attempt to construct classical objects from quantum ones [7].[5] Quantum mechanics as a principle theory is treated, for instance, by J. Bub [8]. For Heisenberg, STR and quantum mechanics are "closed theories". This notion is close to the notion of a principle theory. Within our CQR we have interpreted Heisenberg's "closed theory" as a Wittgenstein rule (W-rule/norm) for measuring reality within the language games of its application [9]. The notion of theory as a W-rule allows us to eliminate the dichotomy between principle theories and constructive theories, although *prima facie* it is closer to the notion of a principle theory. Thus, it seems plausible that Bohr's philosophical views can be interpreted in terms of CQR.

## 2. The anti-fundamentalist ontological interpretation of Bohr's position.

For Bohr, it makes no sense to talk about autonomous quantum systems outside the context of observation, i.e. "interaction" of the macroscopic (classical) instrument and the quantum system. Outside the context of observation, it does not make sense to talk about an observer either. Bohr writes: "The quantum postulate implies that any observation of atomic phenomena will involve an interaction with the agency of observation not to be neglected. Accordingly, an independent reality in the ordinary physical sense can neither be ascribed to the phenomena nor to the agencies of observation" (quoted in [2, p. 11]. Original: Como lecture [10, p. 54]). Therefore, under different experimental conditions one observes different phenomena, which, as we are talking about observation of "the same" quantum system, may be said to be complementary.[6] Bohr's principle of complementarity and the non-physical character of collapse, thus, are closely related with "inseparability" of the observer and the observed system. The "collapse" simply results in a move to a different context of observation (other experimental conditions) and hence to a different phenomenon. Bohr writes: "The essential indivisibility of proper quantum phenomena finds logical expression in the circumstance that any attempt at a well-defined subdivision would require a change in the experimental arrangement that precludes the appearance of the phenomenon itself. Under these conditions, it is not surprising that phenomena observed with different experimental arrangements appear to be contradictory when it is attempted to combine them in a single picture. Such phenomena may appropriately be termed complementary in the sense that they represent equally important aspects of the knowledge obtainable regarding the atomic objects and only together exhaust this knowledge." (Quoted in [2, p. 11], see also [11].) Thus, for Bohr, the "measurement process" is not a spatio-temporal causal process. The so-called "interaction" of the observer and the quantum system, as has been said before, is not a physical interaction. In a sense, the quantum system and the measuring instrument (observer) are not separable from each other (rather than causally interacting with each other). From the point of view of CQR, which here converges with Bohr's position, this inseparability is epistemological, not ontological.[7] But many scholars of Bohr's position consider this to be a neo-Kantian position, whereas CQR rejects transcendentalism. K. Barad, for instance, interpreting Bohr's position, writes: "In my agential realist elaboration of Bohr's account, *apparatuses are the material conditions of possibility and impossibility of mattering*; they enact what matters and what is excluded from mattering" [12, p. 148]. From the CQR point of view, the observer (instruments) plays a logical (grammatical, functional), not a transcendental, role.[8]

---

[2] Some authors see the requirement to separate, during a measurement, the quantum system from the measuring instrument and the environment as a pragmatic point.

[3] For Aristotle, the logical structure of the phenomenon is defined by his categories. In *Metaphysics* he writes: "(…) the appearance is true; not in itself, but for him to whom it appears, and at the time when it appears, and in the way and manner in which it appears" [3, book 4, chapter 6].

[4] De Ronde writes: "A smoky dragon is an irrepresentable meaningless concept which provides a pseudo-picture of a physical situation or process and, consequently, the illusion of understanding. Such inconsistent concepts have no mathematical representation nor posses any operational testability procedure" [5, p. 9].

[5] The distinction between a principle theory and a constructive theory was introduced by A. Einstein. Roughly speaking, the former accounts for the observable phenomena on the basis of general principles, whereas the latter – on the basis of their microscopic internal structure and corresponding causal relationships. For Einstein, STR is a principle theory.

For Bohr, the instrument, if considered as a measuring instrument, i.e. as a classical one, is not entangled with the measured quantum system, and, accordingly, the wave function (exactly) represents the quantum system in an (experimental) context which is described classically (this includes the description of the instrument and the experimental conditions). A clear and not arbitrary distinction between the classical and the quantum is possible, but only in context, and a representation of a quantum system by means of a wave function has a symbolic character (the wave function is not a physical field) [7]. In fact, the description of a context is part of the description of a phenomenon that has its own logic – the "conditions of possibility" (to which quantum theory belongs) – and within which is given (appearing) a quantum "object". In neo-Kantian terms, quantum mechanics (the quantum grammar, the quantum form of life) is the *a priori* condition of the possibility of quantum phenomena; it is implicit in any quantum experiment. But this is a "relative a priori", as there are other *a priori* as well. For example, classical mechanics (its laws) can be treated as the *a priori* condition of the possibility of classical phenomena. The difference between our CQR and (neo)Kantianism is that for CQR theory, its principles are not *a priori* even in the relative sense, but rather are *a posteriori*, rooted in experience (not in the sense of classical empiricism), reality.[9] It makes sense to speak about a theory, its laws and principles as *a priori* conditions of possibility of phenomena only approximately – within the framework of an already established (conceptualized) field of phenomena – a Wittgensteinian form of life, the "grammar" of which is treated as *a priori*.

---

[6] The general structure of quantum mechanics (quantum system) is invariant, context-independent. For example, the Hamiltonian of a quantum system and operators of physical quantities are invariants.

[7] Also, from our CQR point of view, the famous words that the task of physics is not to find out how Nature is but what we can say about Nature, attributed to Bohr, should be understood as an analytic, not a metaphysical, statement and a critique of metaphysical realism.

[8] One can try to interpret Bohr in the spirit of CQR by, so to speak, rooting in reality the neo-Kantian conditions of possibility or relative a priori (constitutive principles). In this way they lose their *a priori* character and become W-rules. Faye, for example, writes: "In emphasizing the necessity of classical concepts for the description of quantum phenomena, Bohr might have been influenced by Kantian-like ideas or neo-Kantianism. But if so, he was a naturalized or a pragmatized Kantian" [1].

Our CQR shares Bohr's position on contextuality of quantum phenomena in the following interpretation by Zinkernagel: "Three important points need to be made regarding this contextuality: 1) When a measurement is performed (that is, when an irreversible recording has been made), then the context changes, and hence the wave function changes. This can formally be seen as a "collapse" of the wave function, with the square quotes indicating that we are not talking about a physical process in which a real wave collapses. 2) The distinction between an epistemic and ontic view of the wave function is sometimes taken to be that between a "representation of an agent's knowledge of the system" and a "representation of the system". In this sense, Bohr's view of the wave function is ontic, for – given an experimental context – the wave function is not agent-relative (i.e. the wave function does not depend on what a particular agent knows about the system but on the experimental arrangement and what in fact has happened to the system, e.g. whether or not the electron has been registered). 3) The experimental context is classically described, which implies that this context is omitted from the quantum description, and thus not represented by means of a wave function. This furthermore implies that no wave function can be ascribed to the measurement apparatus as a whole" [2, p. 12]. For Bohr, the "collapse" of a wave function is a formal (not physical) concept denoting the transition from a superposition of eigenfunctions of a measured physical quantity to its particular eigenfunction.

According to CQR, as for Bohr, quantum measurement is a measurement in context, or a "transition" to a particular context. There is no physical process of transition: we simply find ourselves in this or that context (out of context, the measurement simply has no meaning). In this sense, the measurement problem (reduction/collapse of the wave function, or duality of quantum laws) simply does not arise. CQR also insists that the distinction between the observable system and the observer is categorical (the quantum systems are real, the observers/instruments are ideal), and that quantum ontology is context-sensitive [8]. Remarkably, it is in this direction that Zinkernagel's interpretation of Bohr's position develops. He considers Bohr's point of view to be ontological contextualism: depending on the context, the same "object" can be quantum or classical [2, p. 16]. For example, depending on the context, the atomic nucleus can be interpreted as either a classical or a quantum system. This means that Bohr is a total quantum anti-fundamentalist and not just an epistemological anti-fundamentalist [14].[10] A difference between our CQR and Bohr's view is that the former replaces Bohr's dualism of the quantum and the classical by the categorical one of the real and the ideal.

To this it is appropriate to add the well-known fact, to which Zinkernagel also refers, that for Bohr quantum mechanics is a rational generalization of classical mechanics [2, p. 17]. This conclusion can be best understood within our treatment of classical and quantum mechanics in terms of W-rules (norms), and the transition from classical mechanics to quantum mechanics as a generalization of a W-rule/norm [15]. This approach assumes that in its domain of applicability classical mechanics is a true theory, and classical ontology, in general, is not reduced to quantum ontology (one can speak of this or that local reduction only in this or that context). The items which are being identified within the classical language games, i.e. correct applications of classical concepts, are as real as those being identified within the quantum language games, i.e. correct applications of quantum concepts.

---

[9] Strictly speaking, we accept T. Williamson's view that the distinction between *a priori* and *a posteriori* is superficial and plays no important epistemological role [13].

[10] On the contrary, for example Dorato and some other authors consider Bohr's anti-fundamentalism to refer only to the epistemic level [1; 8].

### 3. Conclusion

In conclusion, we note that Bohr's interpretation of quantum mechanics has been severely criticised by some physicists and philosophers of physics. For example, according to E.T. Jaynes "our present QM formalism is a peculiar mixture describing in part laws of Nature and in part incomplete human information about Nature—all scrambled up together by Bohr into an omelette that nobody has seen how to unscramble" [16, p. 387]. C. Fuchs and some others try to deal with this "omelette" by means of QBism – a variety of quantum Bayesianism. As stated above, Wheeler believed that Bohr operates with pseudo-concepts. A bit closer to truth, in our view, is the point of view of C. De Ronde, according to which Bohr's approach represents "an inconsistent form of anti-realist realism". But that is where De Ronde sees its strength and effectiveness. The critical evaluation of Bohr's position as a "bad philosophy", i.e. a philosophy which is not only wrong, but also blocks the development of physics, is made by D. Deutsch in chapter 12 of his book "The Beginning of Infinity". According to Deutsch, Bohr combines "instrumentalism and studied ambiguity" [6, p. 308]. In our view, the Copenhagen interpretation (in fact its representatives – Bohr, Heisenberg, Pauli, Born, Jordan, von Neumann and others – had their own philosophical views which differed considerably in some respects) breaks with some premises and assumptions of modern philosophy, but not with all of them. This interpretation is considered anti-realist because it is epistemic, i.e. it considers quantum events rather than the "external" reality behind them. Later "realistic" interpretations of quantum mechanics (quantum mechanics with hidden variables, metaphysical many-worlds interpretation, theory of wave function dynamical collapse as a causal process, and others) also remain within the modern philosophy paradigm and, philosophically, have even taken a step back towards metaphysical realism, asserting the existence of an "objective external world". Our CQR is a realist approach without the metaphysics of the otherworldliness. CQR rejects the premises of the (post)modern paradigm [17–20].